
\input phyzzx
\def\ls#1{_{\lower1.5pt\hbox{$\scriptstyle #1$}}}
\def\msusy{M_{\rm SUSY}}
\def\etal{{\it et~al.}}
\def\eg{{\it e.g.}}
\def\ie{{\it i.e.}}
\def\mpl{M_{P}}
\def\wh{\widehat}
\def\vs{{\it vs.}}
\def\hl{h^0}
\def\ha{A^0}
\def\mhl{m_{\hl}}
\def\mha{m_{\ha}}
\def\mw{m_W}
\def\mt{m_t}
\def\tanb{\tan\beta}
\def\SCIPP{\centerline {\it Santa Cruz Institute for Particle Physics}
  \centerline{\it University of California, Santa Cruz, CA 95064}}
\Pubnum={SCIPP 93/22}
\date={July 1993}
\titlepage
\singlespace
\vbox to 4cm{}
\centerline{{\twelvebf The Supersymmetric Top-Ten Lists}
\foot{Work supported in part by the U.S.~Department of Energy.}}
\vskip.8cm
\centerline{\caps Howard E. Haber}
\vskip.1in
\SCIPP
\vskip1.8cm
\vbox{%
\centerline{\bf Abstract}
\vskip4pt
Ten reasons are given why
supersymmetry is the leading candidate for physics
beyond the Standard Model.  Ultimately, the experimental discovery
of supersymmetric particles at future colliders will determine whether
supersymmetry is relevant for TeV scale physics.   The grand hope of
supersymmetry enthusiasts is to connect TeV scale supersymmetry with
Planck scale physics.  The ten most pressing theoretical problems
standing in the way of this goal are briefly described.
}
\vfill
\centerline{Invited Talk presented at the Workshop on Recent Advances
in the Superworld}
\centerline{Houston Advanced Research Center, April 14--16, 1993.}
\vfill
\endpage

\chapter{Introduction}

The organizers of this workshop asked me to give a talk entitled
``Low-Energy Supersymmetry Basics''.   I assume that the
organizers intended this to serve as an
introduction to subsequent talks on supersymmetry in particle
physics.  In this regard, I
thought that it would be useful to
present a collection of the main reasons
we all have attended this workshop.  Namely, why is it that we are all
partisans of supersymmetry, and what are the main theoretical
problems that keep us interested.

\REF\kane{G.L. Kane, UM-TH-93-10 (1993), presented at the
Coral Gables Conference on ``Unified Symmetry in the Small and in the
Large'' (January, 1993), and at the XVIII Rencontre de Moriond, Les
Arcs, France (March, 1993).}
Seven months prior to this workshop, I was attending another workshop,
not all that dissimilar to this one, held in Erice.  During a sumptuous
Italian meal, Gordon Kane made the provocative statement that there
are already at least seven strong experimental indications that
low-energy supersymmetry is a correct description of nature.  A number
of us joined in on the discussion, sometimes challenging Gordy's
assertions and sometimes trying to augment or modify his list.
Since that evening in Erice, Gordy's list has increased to nine
phenomenological indications for low-energy
supersymmetry which now appear in ref.~\kane.

However, this talk
differs from ref.~\kane\ in a number of important respects.  First,
any list that describes the reasons for pursuing supersymmetry should
contain both theoretical arguments as well as phenomenological hints.
In particular, the phenomenological hints are often ambiguous or open
to a variety of interpretations.  That such hints lend their support
to a supersymmetric interpretation is strengthened by the theoretical
motivations for supersymmetry.  Second, I believe that it is important
to highlight the main theoretical challenges for supersymmetric
theories.  The supersymmetric framework is a very ambitious one: it
attempts to connect physics at low energies (the TeV scale and below)
with the ultimate energy scale of fundamental physics---the Planck
scale.  A list of the main unsolved problems of the supersymmetric
approach will illustrate how far we are from achieving this ultimate
goal.

The plan of this talk is as follows.  I will provide two ``top-ten''
lists (hence the use of the plural in the title).  In section 2, I
will give ten reasons why I (and many supersymmetry enthusiasts)
favor supersymmetry.  Here, there is substantial overlap with Gordy's
list; although I do not subscribe to all the phenomenological
hints of ref.~\kane.  In section 3,  I
provide ten challenges for theorists who attempt to use supersymmetry
to connect low-energy physics and the Planck scale.   In some sense,
the second list is the more important one.  The question of whether
low-energy supersymmetry exists (or more specifically, whether
supersymmetry is responsible for the scale of electroweak symmetry
breaking) is ultimately an experimental one.
Yet even if explicit evidence for low-energy supersymmetry is
eventually revealed, the issues addressed in the
second top-ten list will remain.  Of course, one hopes that
additional information such as a detailed supersymmetric particle
spectrum might significantly improve our chances to successfully
address some of the items on the list in section 3.

An addiction of ``top-ten'' lists may be indicative of a night
person (which is true in this case).  However, unlike the lists of David
Letterman, I will not be overly dramatic and save the best bits for
the end.  The order of the points on each list is somewhat arbitrary.
I invite the reader to re-order the items on the list, as well as to
subtract and/or add various items to them.  In the end, I hope that
these lists serve as an overview to the status of supersymmetry in
particle theories today, as well as providing some impetus as to where
we must focus our attention in the future.

\chapter{%
The First Top-Ten List: Why Believe in Supersymmetry?}

\REF\hehtasi{H.E. Haber, SCIPP 92/33 (1993),
to appear in The Proceedings
of the 1992 Theoretical Advanced Study Institute, Boulder, CO,
June 1992.}
\REF\technistuff{E. Farhi and L. Susskind, {\sl Phys. Rep.} {\bf
74} (1981) 277; R.K. Kaul, {\sl Rev. Mod. Phys.} {\bf 55} (1983)
449.}
\REF\preons{I.A. D'Souza and C.S. Kalman, {\it Preons} (World
Scientific, Singapore, 1992).}
\REF\topmode{For a recent review, see M. Lindner, {\sl Int. J.
Mod. Phys.} {\bf A8} (1993) 2167.}
Low-energy supersymmetry is the leading candidate for physics beyond
the Standard Model.  The simplest model of this type is the Minimal
Supersymmetric extension of the Standard Model (MSSM), which is
reviewed in ref.~\hehtasi.
Of course, there are other candidates as well,
including technicolor,\refmark\technistuff\
composite models,\refmark\preons\ models based on effective
four-fermi Lagrangians (\eg, top-mode condensate
models\refmark\topmode), and
perhaps the true model of nature based on physical principles not yet
invented.  But, an informal poll, taken with the help of SPIRES,
indicates that supersymmetry has attracted the most attention of both
theorists and experimentalists.  Here are ten reasons why.

\item{1.} {\sl Supersymmetry is elegant.}
\vskip2.5pt
\REF\mandula{See, \eg, P. West, {\it Introduction to
Supersymmetry and Supergravity} (World Scientific, Singapore,
1990).}
Supersymmetry is a symmetry that associates fermionic and bosonic
degrees of freedom.  It allows one to evade the famous Coleman-Mandula
theorem\refmark\mandula\
which asserted the impossibility of putting together space-%
time symmetries and internal symmetries in a non-trivial way.  The
twentieth century has seen the triumph of gauge symmetries as the
underlying structure of all theories of fundamental forces and
particles.  Supersymmetry is a beautiful generalization of the concept
of continuous symmetries; it would be surprising if nature did not
make use of it.

\item{2.} {\sl Gravity exists.}
\vskip2.5pt
\REF\gravity{P. van Nieuwenhuizen, {\sl Phys. Rep.} {\bf 68}
(1981) 189.}
\REF\superstrings{M.B. Green, J.H. Schwarz and E. Witten, {\it
Superstring Theory} (Cambridge University Press, Cambridge,
England, 1987).}
Supersymmetry may be the link between theories of elementary particles
and a fundamental theory of gravity.  Local supersymmetric theories
necessarily contain gravity.\refmark\gravity\
Moreover, the only consistent quantum
theories that incorporate gravity are superstring theories which possess
supersymmetry at some stage in the theory.\refmark\superstrings\
It is important to
note that this argument by itself
does not set the energy scale at which supersymmetry breaks.  In
particular, it is theoretically conceivable that
supersymmetry is relevant only at the Planck scale ($M_P
\simeq 10^{19}$~GeV) in which case it
would never affect the physics we see at present or future colliders.
\item{3.} {\sl The gauge hierarchy problem.}%
\Ref\suss{E. Gildener, {\sl Phys. Rev.} {\bf D14} (1976) 1667;
S. Weinberg, {\sl Phys. Lett.} {\bf 82B} (1979) 387;
L. Susskind, {\sl Phys. Rep.} {\bf 104} (1984) 181.}
\REF\nilles{H.P.  Nilles, {\sl Phys. Rep.} {\bf 110} (1984) 1.}
\REF\ibanez{For a recent review, see
L.E. Ib\'a\~nez and G.G. Ross, in {\it Perspectives on
Higgs Physics}, edited by G.L. Kane (World Scientific, Singapore,
1993) p.~229.}
\vskip2.5pt

It is very puzzling how the ratio $m_W^2/M_P^2\simeq 10^{-34}$ emerges from
the fundamental Planck scale theory.  Low-energy supersymmetric models
have the potential to solve this problem.  In such theories, the
effective scale of supersymmetry breaking lies below 1 TeV and
provides the connection to the scale of electroweak symmetry breaking.
The most successful model of this type is the radiative symmetry
breaking scenario of minimal supergravity models.\refmark\nilles\
In this picture,
the effective theory at the Planck scale is a globally supersymmetric
model broken by soft-supersymmetry breaking mass terms of order
1 TeV.  The low-energy consequences of such a theory is revealed by using
renormalization group equations (RGEs) with Planck scale boundary
conditions.  One of the Higgs
scalar squared-masses, which is positive at the Planck
scale, is driven negative due to the effects of the large top-quark
Yukawa coupling; this triggers
electroweak symmetry breaking at the required scale.
Note that because the renormalization group evolution is
logarithmic in nature, an exponential hierarchy between the
electroweak scale and the Planck scale can
develop.\refmark\ibanez\

\item{4.} {\sl Naturalness.}%
\Ref\thooft{S. Weinberg, {\sl Phys. Rev.} {\bf D13} (1976) 974;
{\bf D19} (1979) 1277; L. Susskind, {\sl Phys. Rev.} {\bf D20} (1979)
2619; G. 't Hooft, in {\it Recent Developments in Gauge
Theories,} Proceedings of the NATO Advanced Summer Institute,
Cargese, 1979, edited by G. 't~Hooft \etal\ (Plenum, New York,
1980) p.~135.}
\REF\norenorm{M.T. Grisaru, W. Siegel and M. Ro\v cek, {\sl Nucl.
Phys.} {\bf B159} (1979) 429; I. Jack, D.R.T. Jones and P. West,
{\sl Phys. Lett.} {\bf B258} (1991) 382.}
\vskip2.5pt

Despite the simplicity of the Higgs mechanism
in the Standard Model, the existence of
fundamental scalars in field theory is problematical.  If the
electroweak model is embedded in a more fundamental structure
characterized by a much larger energy scale
(\eg, the Planck scale), the Higgs boson
would tend to acquire mass of order the large scale due to
radiative corrections.  Only by adjusting (\ie, ``fine-tuning'')
the parameters of the
Higgs potential ``unnaturally'' can one arrange a large hierarchy
between the Planck scale and the scale of electroweak symmetry breaking.
This requires new physics beyond the Standard Model.
The virtue of the supersymmetric solution to the hierarchy problem is
that it is natural.  That is, the no-renormalization theorems of
supersymmetry\refmark\norenorm\
guarantee that if the hierarchy is established at
tree-level, it is not upset when radiative corrections are included.

\item{5.} {\sl Unification of gauge couplings.}
\vskip2.5pt
\REF\drtj{M.B. Einhorn and D.R.T. Jones, {\sl Nucl. Phys.} {\bf B196}
(1982) 475; W.J. Marciano and G. Senjanovic, {\sl Phys. Rev.}
{\bf D25} (1982) 3092.}
\REF\amaldiold{U. Amaldi \etal, {\sl Phys. Rev.} {\bf D36} (1987) 1385.}
\REF\susygrand{%
U. Amaldi, W. de Boer and H. Furstenau, {\sl Phys. Lett.} {\bf B260}
(1991) 447; U. Amaldi \etal, {\sl Phys. Lett.} {\bf B281} (1992) 374.}
\def\PRL#1&#2&#3&{\sl Phys.~Rev.~Lett.\ \bf #1 \rm (19#2) #3}
\REF\threshes{R. Barbieri and L.J. Hall, \PRL 68&92&752&;
L.J. Hall and U. Sarid, \PRL 70&93&2673&;
P. Langacker and N. Polonsky, {\sl Phys. Rev.} {\bf D47} (1993)
4028; A.E. Faraggi, B. Grinstein, and S. Meshkov, {\sl Phys.
Rev.} {\bf D47} (1993) 5018.}

Consider the three SU(3)$\times$SU(2)$\times$U(1) gauge couplings
($g_1$, $g_2$, and $g_3$)
evaluated at $m\ls Z$.  Now, run these couplings up to the Planck
scale.  In a low-energy supersymmetric model, where the supersymmetry
breaking scale is characterized by $\msusy$, one uses the RGEs of
the Standard Model for energies between $m\ls Z$ and $\msusy$ and
the RGEs of the supersymmetric model for energies between $\msusy$ and
$M_P$.  Do the three gauge coupling constants meet at a single point
(call it $M_X$)?  In the Standard Model, the answer is no.  In the
MSSM, the answer is yes!  Unification occurs at $M_X\simeq 10^{16}$~GeV
for $\msusy\sim 1$~TeV.  This result has been known for some time,%
\refmark{\drtj,\amaldiold}
although a re-analysis by Amaldi and co-workers a few years ago
based on
LEP data caused a great stir in the particle physics community.%
\refmark\susygrand\
Is this result truly significant?

Before being carried
away by all the hype, consider the reactions of the optimist,
the pessimist, and the cynic.  The optimist says: the unification
of coupling constants is the first experimental verification of
the low-energy supersymmetric scenario.  The pessimist says:
the unification of coupling constants only rules out the simplest
GUT extensions of the Standard Model.  It may imply new physics
at any scale between the weak scale and the GUT scale and says
nothing about TeV scale physics.  The cynic says: in the GUT
extension of low-energy supersymmetry, there are three unknown
parameters: $g\ls U$ (the unified coupling constant at the GUT scale),
$M_X$ (the GUT scale or unification point) and $\msusy$.  Thus the
RGEs for $g_1$, $g_2$ and $g_3$ provide three equations and three
unknowns.  A unique solution is essentially guaranteed, so the
unification of coupling constants is no surprise at all.
The optimist clearly overstates the (experimental) case for
supersymmetry.  On the other hand, the pessimist admits that
the unification of couplings implies that the desert hypothesis
of no new physics between the electroweak scale and the GUT scale is
incorrect.  New physics must enter somewhere between $m\ls Z$ and
$M_X$.  This is an exciting result!  Clearly, low-energy supersymmetry
is one possible model for such new physics.  Although there is no
guarantee that the new physics is associated with the TeV scale,
the arguments based on the hierarchy and naturalness problems of the
Standard Model strongly suggest that new TeV scale physics must exist.
The simplest possible scenario would be one in which this TeV scale
physics also accounts for the unification of couplings.  Finally, the
cynic's remarks that the unification of couplings is guaranteed
is technically true (if we ignore the effects of  supersymmetric
thresholds).  However, in solving the RGEs for $\msusy$ and
$M_X$, there was no guarantee that the coupling constant unification
that emerges would be
consistent with sensible values for these parameters.
The fact that such values correspond precisely to the
expected range of a successful grand unified
extension of low-energy supersymmetry
may be more than coincidental and should not be simply dismissed.

\REF\sarid{L.J. Hall and U. Sarid, LBL-32905 (1992).}
Of course, there are numerous complications to the conclusion that
coupling constant
unification is a hint for low-energy supersymmetry.\refmark\threshes\
One must consider the effects of thresholds, both at the low-energy
scale (\eg, the various MSSM particle masses) and at the
high-energy scale (\eg, the superheavy grand unified particle
masses).  Non-renormalizable operators induced at the Planck scale can
also affect the unification of couplings.\refmark\sarid\
Nevertheless, I believe that
the message is clear.  The unification of couplings is a strong hint
for grand unification at scales near $M_P$.  The failure of
coupling constant unification in the Standard Model means that
there is no desert between $m\ls Z$ and $M_X$---new physics at some
intermediate scale must exist.
The fact that coupling constant unification
does occur in the MSSM presents an intriguing clue that the
physics of the desert may have been identified!

\REF\pdecay{P. Langacker, UPR-0539-T (1992), invited talk
given at The Benjamin Franklin Symposium in Celebration of the
Discovery of the Neutrino, Philadelphia, PA, April 29---May 1, 1992.}
\REF\nathsusyproton{R. Arnowitt and P. Nath, {\sl Phys. Rev. Lett.}
{\bf 69} (1992) 725; {\sl Phys. Lett.} {\bf B287} (1992) 89;
NUB-TH-3056/92 (1992);
J. Hisano, H. Murayama and T. Yanagida, Tohoku preprint TU-400 (1992).}
\item{6.} {\sl Proton decay has not yet been observed.}\refmark\pdecay\
\vskip2.5pt
If one accepts the grand unification scenario just discussed,
then one must consider carefully the predictions of proton decay.
It is interesting to note that non-supersymmetric grand unified
models tend to predict proton decay rates that are incompatible
with current experimental bounds (primarily because $M_X$ tends to
lie below $10^{15}$~GeV).
In contrast, in supersymmetric
grand unified models, $M_X$ turns out to be significantly larger,
and the conventional proton decay modes are unobservable.  One must
still check that other decay modes that are induced by new
(dimension-five) operators particular to supersymmetric models
are consistent with present experimental bounds.  This imposes
interesting constraints in some cases
but does not rule out supersymmetric
grand unified models.\refmark\nathsusyproton\
\item{7.} {\sl Relations between third generation quark and lepton masses.}
\vskip2.5pt
\REF\ramondgang{H. Arason \etal, {\sl Phys. Rev.} {\bf D46}
(1992) 3945.}
\REF\twoloop{V. Barger, M.S. Berger and P. Ohmann, {\sl Phys. Rev.}
{\bf D47} (1993) 1093.}
\REF\fermimass{S. Dimopoulos, L.J. Hall and S. Raby,
{\sl Phys. Rev. Lett.} {\bf 68} (1992) 1984; {\sl Phys. Rev.} {\bf D45}
(1992) 4192; {\bf D46} (1992) 4793.}

If one uses Standard Model RGEs and assumes no new physics between the
electroweak scale and $M_X$, then
the prediction of $m_b=m_\tau$ at the unification scale is not
compatible with low-energy data.\refmark\ramondgang\
However, the relation
$m_b=m_\tau$ at $M_X$ is still viable in supersymmetric
grand unified models.\refmark\twoloop\
Relations among other quark and lepton masses
require a more complicated structure at the grand unification scale.
Supersymmetric models
can accommodate such a structure, although it is not clear whether
this constitutes a real hint for low-energy supersymmetry.%
\refmark\fermimass\

\REF\witten{E. Witten, {\sl Nucl. Phys.} {\bf B258} (1985) 75.}
\REF\yukawas{See, \eg, G.G. Ross,
in {\it The Santa Fe TASI-87}, Proceedings of
the 1987 Theoretical Advanced Study Institute, Santa Fe, NM,
edited by R. Slansky and G. West (World Scientific, Singapore, 1988)
p.~628; in {\it Particles and Fields---3}, Proceedings of the 3rd
Banff Summer Institute on Particles and Fields, Banff, Alberta,
August 14---27, 1988, edited by A.N. Kamal and F.C. Khanna
(World Scientific, Singapore, 1989) p.~223.}
The result $m_b=m_\tau$ at $M_X$ may be a less compelling clue
than the unification of gauge coupling constants.  For example,
in some string models, unification of gauge couplings can occur
without grand unification, whereas the Yukawa couplings depend
in part on the structure of the compactification manifold
and do not necessarily satisfy standard
unification relations.\refmark{\witten,\yukawas}

\item{8.} {\sl The existence of cold dark matter.}
\vskip2.5pt
\REF\turner{E.W. Kolb and M.S. Turner, {\it The Early Universe}
(Addison-Wesley Publishing Company, Reading, MA, 1990);
T. Padmanabhan, {\it Structure Formation in the Universe}
(Cambridge University Press, Cambridge, England, 1993).}
\REF\dark{S. Kelley, J.L. Lopez, D.V. Nanopoulos, H. Pois and K. Yuan,
 {\sl Phys. Rev.} {\bf D47} (1993) 2461;
 R.G. Roberts and L. Roszkowski, {\sl Phys. Lett.} {\bf B309}
(1993) 337.}
\REF\darkharc{K. Yuan, contribution to these Proceedings.}

Most theoretical cosmologists believe that
the ratio of the matter density in the
universe to the critical density, $\Omega\equiv\rho/\rho_c=1$.
This result follows from theories of inflation, and there are some
observational hints that also support this conclusion.\refmark\turner\
But, the baryonic matter density cannot contribute more than
$0.2\rho_c$, which strongly suggests the existence of dark matter
making up a significant portion of the total
matter density of the universe.
The precise nature of the dark matter has been much debated in the
astrophysical community.  Evidence based on theories of galaxy
formation and the fluctuations of the microwave background radiation
suggest that a substantial fraction of the dark matter is likely to be
``cold''.  The lightest supersymmetric particle (LSP) is an ideal
candidate
for cold dark matter.  Ranges of MSSM parameter space exist where
the primordial abundance of the LSP provides exactly the right amount
of ``missing mass'' to reach the critical closure density.\refmark\dark\
See ref.~\darkharc\ for further details.

\item{9.} {\sl Precision electroweak measurements at LEP show no deviation
{}from the Standard Model.}
\vskip2.5pt
\REF\tatsu{M.E. Peskin and T. Takeuchi, {\sl Phys. Rev. Lett.} {\bf 65}
(1990) 964; {\sl Phys. Rev.} {\bf D46} (1992) 381.}
\REF\otherst{G. Altarelli and R. Barbieri, {\sl Phys. Lett.} {\bf B253}
(1990) 161.}
\REF\stulimits{P. Langacker, in {\it Electroweak Physics Beyond
the Standard Model}, Proceedings of the International Workshop on
Electroweak Interactions Beyond the Standard Model, Valencia,
Spain, October 2---5, 1991, edited by J.F.W. Valle and J. Velasco
(World Scientific, Singapore, 1992) p.~75.}
\REF\holdom{%
B. Holdom and J. Terning, {\sl Phys. Lett.} {\bf B247} (1990) 88;
M. Golden and L. Randall, {\sl Nucl. Phys.} {\bf B361} (1991) 3.}
\REF\barbieri{R. Barbieri, M. Frigeni, F. Giuliani, and H.E. Haber,
\sl Nucl. Phys. {\bf B341}, \rm 309 (1990).}
\REF\ericetwo{H.E. Haber, SCIPP 93/06 (1993), to appear
in the Proceedings of the 23rd Workshop of the INFN Eloisatron
Project, ``Properties of Supersymmetric Particles'', Erice, Italy,
September 28---October 4, 1992.}

Suppose that the origin of the electroweak scale
lies with new physics beyond the Standard Model.  By the naturalness
arguments of section 2.4, the energy scale
at which this new physics enters must not lie much above 1 TeV.
One must then check that the effects of virtual new heavy particle
exchange
is not in conflict with the precision electroweak measurements at LEP,
which at present show no evidence of departures from the Standard
Model.  One might be tempted to conclude that the effects of new
heavy physics should decouple from LEP observables.  However, this
is not always true, since violation of decoupling can occur in
spontaneously broken gauge theories.  In many cases, the
effects of new heavy particles on electroweak radiative corrections
can be neatly summarized by three different combinations of vector
boson self-energies.\refmark{\tatsu,\otherst}
(These are called oblique radiative corrections.)
$T$ is proportional to the shift
in the $\rho$-parameter and $S$ counts the number of very massive
degenerate chiral weak multiplets.  A third parameter, $U$, also enters
although it is typically smaller than $S$ and $T$.

\FIG\fone{%
The contribution to the $S$ and $T$ parameters from the
neutralino and chargino sector of the MSSM as a function
of $\mu$ for $\tan\beta=2$.  The four curves shown correspond
to $M=50$, $250$, $500$ and $1000$~GeV [with $M_2\equiv M$ and
$M_1=(5g^{\prime 2}/3g^2)M$].  In (a), curves in the region
of $|\mu|\leq100$~GeV are not shown, since in this region of
parameter space the light chargino mass is less than
of order $m\ls Z$.  In (b), $T$ is related to the
$\rho$ parameter via $\delta\rho=\alpha\delta T$
which is an experimental observable over the entire mass parameter
region.  Taken from ref.~\ericetwo.}
Recent LEP
measurements show that the contributions to $S$, $T$ and $U$ from
new physics beyond the Standard Model must be less than
1.\refmark\stulimits\
This is
not a trivial constraint.  For example, it has been shown that
$S$ can be reliably estimated in a class of technicolor models.%
\refmark{\tatsu,\holdom}
New heavy technifermion doublets do not decouple from $S$, so
precision electroweak measurements can potentially rule out such
models.   In contrast, supersymmetric models have the property that
their contributions to $S$, $T$ and $U$ precisely decouple in the
limit of large supersymmetry breaking scale, $\msusy$.
Still, one must check the coefficient of
the leading terms to determine
the numerical importance of the supersymmetric
contributions.\refmark{\barbieri,\ericetwo}
I have computed the contributions of the various
supersymmetric sectors to $S$, $T$ and $U$ as a function of the MSSM
parameters.\refmark\ericetwo\  An example of these results is shown
in fig.~\fone.
I find that once supersymmetric particle masses all
become larger than about $150$~GeV, the effects of supersymmetry on
the oblique radiative corrections become negligible.
Thus, the non-observation of deviations from the Standard Model at LEP
is easily compatible with supersymmetric extensions of the Standard
Model, in contrast to other excursions beyond the Standard Model mentioned
above.

\REF\beesgam{%
A. Buras, P. Krawczyk, M.E. Lautenbacher and C. Salazar,
{\sl Nucl. Phys.} {\bf B337} (1990) 284;
S. Bertolini, F. Borzumati, A. Masiero and G. Ridolfi,
{\sl Nucl. Phys.} {\bf B353} (1991) 591;
J.L. Hewett, {\sl Phys. Rev. Lett.} {\bf 70} (1993) 1045;
V. Barger, M.S. Berger, and R.J.N. Phillips, {\sl Phys. Rev. Lett.}
{\bf 70} (1993) 1368;
R. Barbieri and G.F. Giudice, {\sl Phys. Lett.} {\bf B309} (1993)
86.}
\REF\cleolims{R. Ammar \etal\ [CLEO Collaboration] CLNS-93-1212
(1993).}

To be complete, it is important to note that some non-oblique
radiative corrections (\eg, vertex corrections) can arise in
supersymmetric models that are observable.  Perhaps the most interesting
example of this type is $b\to s\gamma$ which is induced in the Standard
Model at one-loop.\refmark\beesgam\
In the MSSM, new contributions enter which could
alter the Standard Model prediction.  Recent bounds from
CLEO\refmark\cleolims\
can already place interesting limits on the MSSM parameter space.
Nevertheless, in the limit of large supersymmetric masses, these
contributions vanish as well.  Thus, it is likely that supersymmetric
particles, if they exist, will be discovered by direct production at
future colliders before their virtual effects are uncovered.
\REF\lepsearch{See, \eg, D.~Decamp \etal\ [ALEPH Collaboration],
{\sl Phys. Rep.} {\bf 216} (1992) 253.}
\REF\hhg{See chapter 4 of
J.F. Gunion, H.E. Haber, G.L. Kane and S. Dawson, {\it
The Higgs Hunter's Guide} (Addison-Wesley Publishing Company,
Reading, MA, 1990).}
\REF\hhprl{H.E. Haber and R. Hempfling, {\sl Phys. Rev. Lett.} {\bf 66}
(1991) 1815; Y.
Okada, M. Yamaguchi and T. Yanagida, {\sl Prog. Theor. Phys.} {\bf 85}
(1991) 1; J. Ellis, G. Ridolfi and F. Zwirner, {\sl Phys. Lett.}
{\bf B257} (1991) 83; {\bf B262} (1991) 477.}
\REF\onehiggsrge{Y.
Okada, M. Yamaguchi and T. Yanagida, {\sl Phys. Lett.} {\bf B262}
(1991) 54; R. Barbieri, M. Frigeni,
and F. Caravaglios, {\sl Phys. Lett.} {\bf B258} (1991) 167;
J.R. Espinosa and M. Quiros, {\sl Phys. Lett.} {\bf B267}
(1991) 27.}
\REF\moreradmssm{%
R. Barbieri and M. Frigeni, {\sl Phys. Lett.} {\bf B258} (1991) 395;
A. Yamada, {\sl Phys. Lett.} {\bf B263} (1991) 233.}
\REF\berz{A. Brignole, J. Ellis, G. Ridolfi and F. Zwirner,
{\sl Phys. Lett.} {\bf B271} (1991) 123 [E: {\bf B273} (1991) 550].}
\REF\pokorski{
P.H. Chankowski, S. Pokorski and J. Rosiek, {\sl Phys. Lett.}
{\bf B274} (1992) 191; {\bf B281} (1992) 100; MPI-Ph/92-117 and
DFPD 93/TH/13 (1993).}
\REF\brignole{A. Brignole, {\sl Phys. Lett.} {\bf B277} (1992) 313.}
\REF\diaz{M.A. Diaz and H.E. Haber, {\sl Phys. Rev.} {\bf D45} (1992)
4246.}
\REF\berkeley{D.M. Pierce, A. Papadopoulos, and S. Johnson,
{\sl Phys. Rev. Lett.} {\bf 68} (1992) 3678.}
\REF\sasaki{K. Sasaki, M. Carena and C.E.M. Wagner, {\sl Nucl. Phys.}
{\bf B381} (1992) 66.}
\REF\andrea{A. Brignole, {\sl Phys. Lett.} {\bf B281} (1992) 284.}
\REF\diaztwo{M.A. Diaz and H.E. Haber, {\sl Phys.
Rev.} {\bf D46} (1992) 3086.}
\REF\llog{H.E. Haber and R. Hempfling, SCIPP 91/33 (1992),
{\sl Phys. Rev.} {\bf D49} (1993) in press.}
\REF\alephnew{D. Buskulic \etal\ [ALEPH Collaboration], CERN-%
PPE/93-40 (1993).}
\REF\cmpp{N. Cabibbo, L. Maiani, G. Parisi, R. Petronzio,
{\sl Nucl.~Phys.} {\bf B158} (1979) 295.}
\REF\habersher{H.E. Haber and M. Sher, {\sl Phys. Rev.} {\bf D35}
(1987) 2206; M. Drees, {\sl Phys. Rev.} {\bf D35} (1987) 2910;
{\sl Int. J. Mod. Phys.} {\bf A4} (1989) 3635;
J. Ellis, J.F. Gunion, H.E. Haber, L. Roszkowski, and
F. Zwirner, {\sl Phys. Rev.} {\bf D39} (1989) 844;
J.R. Espinosa and M. Quiros, {\sl Phys. Lett.} {\bf B279} (1992) 92;
{\bf B302} (1993) 51.}
\REF\kanehiggs{G.L. Kane, C. Kolda and J.D. Wells, {\sl Phys.
Rev. Lett.} {\bf 70} (1993) 268.}
\REF\precise{J. Ellis, G.L. Fogli
and E. Lisi, {\sl Phys. Lett.} {\bf B279} (1992) 169;
{\sl Phys. Lett.} {\bf B285} (1992) 238;
{\sl Phys. Lett.} {\bf B286} (1992) 85; {\sl Nucl. Phys.} {\bf B393}
(1993) 3;
F. del Aguila, M. Martinez and M. Quiros
{\sl Nucl. Phys.} {\bf B381} (1992) 451.}

\item{10.} {\sl  LEP has not discovered the Higgs
boson.}\refmark\lepsearch\

\vskip2.5pt
The tenth reason is admittedly given with a little tongue in cheek.
(I am sure that technicolor enthusiasts would place this point prominently
on their top-ten list.)  Nevertheless, low-energy supersymmetry
leads to important constraints  on the Higgs sector.  The experimental
discovery or absence of the Higgs boson in future experiments will have a
significant impact on the validity of low-energy supersymmetry.
Consider first the MSSM.  The Higgs sector of the MSSM is a
constrained two-Higgs-doublet model, in which all quartic Higgs self-%
couplings are given in terms of the gauge couplings.\refmark\hhg\
This means that
at least one physical Higgs boson of the model cannot be arbitrarily
heavy.  It is easy to show that the lightest CP-even Higgs scalar
satisfies the following tree-level bound:
$0\leq\mhl\leq m\ls Z|\cos2\beta|$,
where $\tanb$ is the ratio of Higgs vacuum expectation values.   In
light of this result, it seems that the LEP non-discovery of the Higgs
is somewhat of an embarrassment for the MSSM.

\FIG\ftwo{%
RGE-improved Higgs mass $\mhl$ as a function
of $\tanb$ for (a) $\mt = 150$ GeV and (b) $\mt = 200$ GeV.  Various
curves correspond to $\mha = 0,~20,~50,~100$ and $300$ GeV as labeled in
the figure. All $A$-parameters and $\mu$ are set equal to zero.
The light CP-even Higgs mass varies very weakly with $\mha$
for $\mha>300$ GeV.  Taken from ref.~\llog.}
This perception changed dramatically a few years ago when it was
realized that
both the lower and upper bounds on the lightest Higgs mass in the MSSM
are significantly affected by radiative corrections.  This is mainly
a result of the incomplete cancellation between top quark and top-squark
loop corrections to the Higgs two-point function.  For example, the
most significant one-loop
radiative correction to the neutral CP-even Higgs
masses grows as $m_t^4\ln(M_{\tilde t}^2/\mt^2)$.
Since these effects were first uncovered independently by three
groups,\refmark\hhprl\
there have been many papers in the literature examining the impact
of the radiative corrections on the MSSM Higgs sector.%
\refmark{\onehiggsrge-\llog}  Here, I shall quote only two results.
First, if $\tanb=1$, then the tree-level prediction for the light
CP-even Higgs mass is $\mhl=0$.  In this case, the physical mass of
the $\hl$ is entirely due to radiative corrections.
Marco Diaz and I have performed an exact one-loop
calculation of the neutral Higgs masses for values of $\tanb$ near
1.\refmark\diaztwo\
We confirmed that for $\tanb$ near 1, large radiative
corrections to the light Higgs mass can easily push $\mhl$ to values
beyond the current LEP experimental lower bound of 60 GeV.
Thus, the possibility that the light Higgs mass is due entirely to
radiative corrections is not ruled out!  Second, in fig.~\ftwo, I
show the predicted value of $\mhl$ \vs\ $\tanb$ for
various values of the CP-odd Higgs mass, $\mha$.   Two graphs are shown
corresponding to $\mt=150$~GeV and 200~GeV, respectively.  For values
of $\mha>150$~GeV (assuming characteristic supersymmetric particle
masses of order 1~TeV),
one sees that the predicted value for $\mhl$ lies
above the current LEP Higgs mass bound.

The fact that a significant region of MSSM parameter space leads to a
predicted value of $\mhl>60$~GeV is a consequence of enhanced radiative
corrections driven by a large top-quark mass.
To put it another way, given the large value of $m_t$ suggested by
LEP precision electroweak experiments,\refmark\alephnew\
the most probable MSSM parameters would put the light Higgs boson
outside the reach of LEP-I.

So far, the above discussion has focused on the MSSM.  What about
non-minimal models of low-energy supersymmetry?
In such models, the main new feature
is the possibility of Higgs-self couplings that are not related
to the gauge couplings.  Top-quark mass enhanced radiative corrections
would still tend to raise the tree-level value of the lightest Higgs
boson.  But, the upper Higgs mass limit is seemingly unconstrained,
since it depends on a new unknown parameter.  In this case, it is
tempting to make use of the observation of gauge coupling constant
unification to conclude that {\it no new physics enters between the
TeV scale and $M_X$}.  If this is true, then it seems likely that all
couplings of the model remain perturbative below $M_X$.  In the case
of the Standard Model, the
Higgs squared-mass is proportional to the Higgs self-coupling.
The renormalization group scaling of this coupling indicates that
the Landau pole would be reached below the Planck scale (indicating
that new physics must enter) if the Higgs self-coupling at the
electroweak scale lies above a certain value.  This result translates
into the Higgs mass bound: $m_H\lsim 175$~GeV.\refmark\cmpp\
In non-minimal low-energy
supersymmetric models, new Higgs self-coupling parameters enter
which must be bounded in the same way as in the Standard Model.%
\refmark{\habersher,\kanehiggs}
This yields a similar Higgs mass upper bound, which according to
ref.~\kanehiggs\ is around 150~GeV.

Thus, if no Higgs boson is found below 150~GeV (and assuming that
such a Higgs scalar is not unexpectedly difficult to detect by the
standard experimental techniques), one would have to be prepared to
either give up on low-energy supersymmetry or abandon the concept
of the desert between 1~TeV and the Planck scale.  Is there any
experimental hint that the Higgs mass might be light (\ie, of order
$m\ls Z$ rather than, say, 1~TeV)?  Without a good measurement of the
top-quark mass, the precision electroweak data from LEP is not
accurate enough for one to reach any conclusion on the Higgs mass.
Nevertheless, there are intriguing hints from some of the theoretical
analyses of LEP data that give a weak preference for light Higgs mass
values.\refmark\precise\
If such an indication is confirmed, it could provide yet another
confirmation of the expectations of low-energy supersymmetry.
Of course, the discovery of the Higgs boson at LEP-II would give a
much larger boost to low-energy supersymmetry.  Based on the Higgs
mass calculations quoted above, I give LEP-II about a 50-50
chance for a Higgs discovery if they can extend their search up
to $m\ls Z$.  This may be the best hint for low-energy supersymmetry
prior to turning on the supercolliders.

\chapter{%
The Second Top-Ten List: Challenges to a Supersymmetric Theory}
\vskip-\headskip
\leftline{\bf of Particle Physics.}
\vskip\headskip

The primary motivation for supersymmetry is that it has the potential
for providing a consistent,
natural embedding of the Standard Model of particle
physics in a more fundamental theory whose natural scale is $\mpl$.
The unification of coupling constants discussed in section 2.5 provides
the strongest hint that one can extrapolate from the TeV scale
all the way up to energies near $\mpl$.  However, a fundamental
supersymmetric theory of particles remains an elusive goal.  Many
theorists insist that the fundamental supersymmetric theory can be
truly understood only in the context of superstring theory.  This is a
very ambitious point of view which proposes that superstring theory is
the ``theory of everything''.  For example, in such a framework,
one could in principle derive
the effective low-energy broken-supergravity model that emerges at the
Planck scale.  If this is your point of view, then perhaps you should
replace the following list with the top-ten list of the
outstanding problems in string theory and string model building.  Such
a list would contain questions such as: (i) what is the correct string
vacuum? (ii) how does one compute the effective Planck-scale broken-%
supergravity model parameters from string theory? (iii) \etc\
These questions lie beyond
the scope of this talk (and my expertise), and I refer you to the
string talks of this workshop.  Nevertheless, it is certainly
worthwhile to contemplate the solutions to the questions posed in the list
below in the context of string theory.

Here are
ten theoretical problems that must be overcome on the way to
constructing a successful supersymmetric theory of particle physics
{}from the TeV scale to the Planck scale.

\item{1.} {\sl The origin of supersymmetry breaking.}

\vskip2.5pt
\REF\hlw{See, \eg, L. Hall, J. Lykken and S. Weinberg, {\sl Phys. Rev.}
{\bf D27} (1983) 2359; S.K. Soni and H.A. Weldon, {\sl Phys. Lett.}
{\bf 126B} (1983) 215.}
\REF\gauginocondensate{%
H.-P. Nilles, {\sl Int. J. Mod. Phys.} {\bf A5} (1990) 4199.}
\REF\stringsusy{L.E. Ib\'a\~nez and D. Lust, {\sl Nucl. Phys.} {\bf B382}
(1992) 305; V.S. Kaplunovsky and J. Louis, {\sl Phys. Lett.} {\bf
B306} (1993) 269.}
The origin of supersymmetry breaking is one of the
most pressing theoretical problem in fundamental theories of
supersymmetry.  Here, I shall only briefly outline
the most common scenario for producing low-energy supersymmetry from a
more fundamental broken supergravity model.  This scenario has been
called the {\it hidden sector}$\,$ scenario.\refmark\hlw\
In this scenario, one posits
two sectors of fields.  One sector (called the ``visible'' sector)
contains all the fields of the
Standard Model (and perhaps additional heavy
fields in a grand unified model
of the strong and electroweak forces).  A second ``hidden'' sector
contains fields which lead to the breaking of supersymmetry at some
large scale $\Lambda_{\rm SUSY}$.  One assumes that none of the fields
in the hidden sector carry quantum numbers of the visible sector.
Thus, the two sectors are nearly decoupled; they
communicate only by weak gravitational interactions.  Thus, the
visible sector only finds out about supersymmetry breaking through its
very weak gravitational
couplings to the hidden sector.  In the visible sector, the
effective scale of supersymmetry breaking (denoted by $\msusy$)
is therefore much smaller than $\Lambda_{\rm SUSY}$.  A typical result
is
$$
\msusy\simeq{\Lambda_{\rm SUSY}^n \over M_P^{n-1}}\,,\eqn\msusybreak
$$
depending on the mechanism for supersymmetry breaking in the hidden
sector.  Two popular models for the breaking mechanism are the Polonyi
model (where $n=2$) based on $F$-type breaking in the hidden sector,
and gaugino condensate models\refmark\gauginocondensate\
(where $n=3$).  In both cases, $\Lambda_{\rm SUSY}$
can be quite large, above $10^{10}$~GeV, while still producing
$\msusy$ of order 1 TeV or less.

In these scenarios,
supersymmetry has the potential for solving the hierarchy and
naturalness problems described sections in 2.3 and 2.4.
However, at this point, we
have only scenarios rather than realistic models.  The gaugino
condensate model is indicative of the difficulty in constructing a
realistic and viable fundamental
model of supersymmetry breaking.  It suggests
that the origin of supersymmetry breaking is probably nonperturbative.
As a result, reliable calculations are difficult.  This
situation is somewhat reminiscent of the status of technicolor
approaches to electroweak symmetry breaking; \ie, a number of
scenarios have been advanced, but no standard model of technicolor
exists.  In this regard, I would like to make a plea to the advocates
of technicolor and related strong interaction approaches to
electroweak symmetry breaking.  Lend us your skills of
nonperturbative analysis and help us to unravel the secrets of
the fundamental origin of supersymmetry breaking!

\REF\vadim{V. Kaplunovsky, contribution to these Proceedings.}
In string models, supersymmetry breaking should emerge as a
consequence of the dynamics of the model.  If one were able to
successfully solve the string theory and determine the correct
vacuum, one would in principle have the tools for determining the
``low-energy'' effective broken supergravity model at
the Planck scale.  The soft-supersymmetry breaking parameters would
then be computable, and would serve as boundary conditions for
renormalization group evolution down to the electroweak scale.
Recently, there have been some attempts to explore model-independent
features of the soft-supersymmetry breaking terms that emerge from
string theory.\refmark\stringsusy\ Kaplunovsky will summarize some
of these results later in this workshop.\refmark\vadim\

\item{2.} {\sl The cosmological constant problem.}
\vskip2.5pt

\REF\cosmoconstant{S. Weinberg, {\sl Rev. Mod. Phys.}
{\bf 61} (1989) 1.}
Even if one is successful in making use of supersymmetry to solve the
gauge hierarchy problem with no fine-tuning,
there is one unsolved fine-tuning problem
which remains, called the cosmological constant
problem.\refmark\cosmoconstant\ Once we take
gravity into account in particle theory, the vacuum energy
density ($\Lambda_0$)
is a physical quantity which in principle is calculable in a
fundamental theory of gravity.  Theoretically, the vacuum energy
density is naively expected to be of order $\mpl^4$, but this is
not our universe.  (A universe with such a large vacuum energy density
would have a lifetime of order the Planck time, $\hbar/M_P c^2\simeq
10^{-43}$~sec!)  Thus, based on the fact that the universe has
endured over 10 billion years (and looks very flat at large scales),
$\Lambda_0/\mpl^4<10^{-121}$.  This is the mother of all fine-tuning
and naturalness problems!

The extent of this fine-tuning problem is
slightly alleviated in broken supersymmetric models.  But at best,
$\mpl^4$ is replaced by $\msusy^4$.  Most theorists simply put this
question aside, perhaps to be solved at an undetermined future time.
However, it is not clear that this is justified.  One could ask
whether one should accept the theoretical motivation of low-energy
supersymmetry to solve the gauge hierarchy and fine-tuning problems
while ignoring the most severe fine-tuning problem of them all.
Perhaps a solution of the cosmological constant problem will
automatically solve all other fine-tuning problems by some
presently unknown theoretical mechanism, thus rendering supersymmetry
unnecessary.  Although it is difficult to argue against such a
proposition, the fact that supersymmetry seems to significantly reduce the
severity of the cosmological constant problem (as indicated above)
may be a hint that supersymmetry will play a key role in its eventual
solution.

\REF\casas{B. de Carlos, J.A. Casas and C. Mu\~noz, {\sl Nucl. Phys.}
{\bf B399} (1993) 623.}
In the absence of a solution to the cosmological constant problem,
one must simply be prepared to accept for now
the required fine-tuning in
models that incorporates both particle physics and gravity.
In models of spontaneously-broken
supergravity, there is some freedom that allows a
fine-tuning of parameters to set the cosmological constant to zero.
All model builders must do this in order to have a
theoretically consistent framework.  In some string models, the
cosmological constant cannot be adjusted by hand, rather it is
fixed by the theory.  For example, in the models of ref.~\casas, the
cosmological constant is of order $\msusy^4$.  While this is
certainly an improvement over the natural value of $\mpl^4$, it is
difficult to ascertain whether one can make sense of such models
by pretending that the cosmological constant is zero.

\REF\mxmpl{I. Antoniadis, J. Ellis, R. Lacaze and D.V. Nanopoulos,
{\sl Phys. Lett.} {\bf B268} (1991) 188; I. Antoniadis, J. Ellis,
S. Kelley and D.V. Nanopoulos, {\sl Phys. Lett.} {\bf B272}
(1991) 31; S. Kelley, J.L. Lopez and D.V. Nanopoulos, {\sl Phys.
Lett.} {\bf B278} (1992) 140.}

\item{3.} {\sl The origin of the unification scale ($M_X$).}

\vskip2.5pt
In section 2.5, we saw that the unification of gauge coupling constants
takes place at $M_X\simeq 10^{16}$~GeV.  This implies that
$M_X\neq\mpl$.  How is the scale $M_X$ generated?  Perhaps threshold
effects of super-heavy particles are sufficient to push $M_X$ toward
$\mpl$ such that the two scales are not distinct.  In some superstring
theories, the unification of coupling constants is predicted to occur
at $\mpl$, so the fact that $M_X\neq\mpl$ is somewhat
problematical.  One possible way around this problem is to
add extra multiplets to the theory to
delay the unification to $\mpl$; some examples can be found in
ref.~\mxmpl.

\item{4.} {\sl The gauge hierarchy and tree-level fine-tuning problem.}

\vskip2.5pt
\REF\grin{H. Georgi, {\sl Phys. Lett.} {\bf B108} (1982) 283;
A. Masiero, D.V. Nanopoulos, K. Tamvakis, and T. Yanagida, {\sl
Phys. Lett.} {\bf 115B} (1982) 380; B. Grinstein, {\sl Nucl.
Phys.} {\bf B206} (1982) 387; S.-C. Chao, {\sl Nucl. Phys.} {\bf
B256} (1985) 705.}
\REF\othergrin{A.A. Anselm and A.A. Johansen, {\sl Phys. Lett.}
{\bf B200} (1988) 331; G.R. Dvali, {\sl Phys. Lett.} {\bf B287}
(1992) 101.}
\REF\dine{I thank Michael Dine for stressing this point to me.}
One of the theoretical motivations of supersymmetry is to solve the
gauge hierarchy and naturalness problems, as discussed in sections 2.3
and 2.4.
In supersymmetric grand unified models, the ratio $\mw^2/M_X^2\simeq
10^{-28}$ is stable under radiative corrections.  This
follows from the
supersymmetric no-renormalization theorems,\refmark\norenorm\
which imply
that the parameters of the superpotential (where the above ratio is
set) are not renormalized.  Still, one can ask: where does such a small
number arise in the first place?  For example, in supersymmetric
SU(5), this small number must be inserted into the theory (\ie, in the
superpotential) by hand at tree-level.  Specifically, one must fine-%
tune tree-level parameters of the theory to an accuracy of one part in
$10^{28}$, in order that the uncolored doublet Higgs fields remain
light (of order $\mw$) while the color triplet Higgs fields are
superheavy (with masses of order $M_X$).   This is the famous doublet-%
triplet mass splitting problem which is shared by most grand unified
theories.  In ordinary SU(5), this hierarchy is unstable under
radiative corrections.  In supersymmetric SU(5), one can ``set it and
forget it'', but this is not a desirable attribute of a fundamental
theory.  Possible solutions to the tree-level fine-tuning problem
do exist.  One such example is the missing-partner
mechanism;\refmark\grin\  other solutions have also been
proposed.\refmark\othergrin\
However, having eliminated tree-level fine-tuning does not
guarantee that the gauge hierarchy problem has been solved.  One must
check that higher dimensional operators (which are suppressed by
inverse powers of the Planck scale) do not
re-introduce fine-tuning (which may be less severe than the original
fine-tuning, but may still be too large for comfort)
in order to maintain the required gauge hierarchy.\refmark\dine\
String theory also provides another mechanism for generating
(approximately)
massless Higgs doublets, in models where no formal grand unification
occurs.\refmark\witten

\item{5.} {\sl The $\mu$-problem.}%
\Ref\muprob{For a recent discussion, see J.A. Casas and C.
Mu\~noz, {\sl Phys. Lett.} {\bf B306} (1993) 288.}
\vskip2.5pt

\REF\bagger{This point has recently been re-examined in
J. Bagger and E. Poppitz, Johns Hopkins preprint (1993).}
A low-energy supersymmetric model is specified by its superpotential
and collection of soft-supersymmetry breaking terms.  The latter are
dimension two or three terms with coefficients with units of mass to
the appropriate power.  The scale of the soft-supersymmetry-breaking
terms is the electroweak scale; the origin of this scale is tied to
the mechanism of supersymmetry-breaking, discussed in section 3.1.
But what about the terms in the superpotential with units of mass?  For
example, the superpotential of the MSSM contains the term $\mu \widehat
H_1 \widehat H_2$
(where the $\widehat H_i$ are the two Higgs superfields),
The parameter $\mu$ has dimensions of mass, which must be no larger than
about 1 TeV in order to preserve the naturalness of the electroweak
theory.  What is the origin of the scale $\mu$?  There is danger that
in the fundamental theory at the Planck scale, $\mu$ could be
generated with a value of order $\mpl$.  Perhaps the most natural
solution is to demand that only dimensionless parameters in the
superpotential can be nonzero.  However, this is not acceptable in the
case of the MSSM, since by setting $\mu=0$, the theory would have a
Peccei-Quinn symmetry, leading to a weak scale axion which is
experimentally untenable.\foot{In the standard low-energy supergravity
approach, if $\mu=0$, then so is the soft-supersymmetry breaking term
$m_{12}^2\equiv B\mu$ which is the coefficient of $H_1 H_2$ in the
scalar potential.}  Other solutions to the $\mu$-problem have been
proposed.  One solution is to add a singlet superfield $\widehat N$
to the theory and eliminate the $\mu$-term in the superpotential in
favor of the term $\widehat H_1\widehat H_2\widehat N$.  Then, $\mu$
would be generated when $\wh N$ acquires a vacuum expectation value.
However, singlet superfields are dangerous in that they can destroy
the gauge hierarchy.\refmark\bagger\    A more
natural mechanism is one in which $\mu=0$ initially, but
a non-zero value of $\mu$ of order 1~TeV (or less) is generated
when supersymmetry-breaking effects are taken into account.

\item{6.} {\sl The gravitino problem.}
\vskip2.5pt
\REF\gravitino{S. Weinberg, {\sl Phys. Rev. Lett.} {\bf 48}
(1982) 1303.}
\REF\moregravitino{J. Ellis, A.D. Linde and D.V. Nanopoulos, {\sl Phys.
Lett.} {\bf 118B} (1982) 59;
S. Dimopoulos and S. Raby, {\sl Nucl. Phys.} {\bf B219} (1983)
479; L.M. Krauss, {\sl Nucl. Phys.} {\bf B227} (1983) 556; J.
Ellis, J.E. Kim and D.V. Nanopoulos, {\sl Phys. Lett.} {\bf 145B}
(1984) 181; B.A. Ovrut, {\sl Phys. Lett.} {\bf 147B} (1984) 263.}
\REF\baryons{For a recent review, see A.G. Cohen, D.B. Kaplan and
A.E. Nelson, UCSD-PTH-93-02 (1993).}
\REF\nelson{A.G. Cohen and A.E. Nelson, {\sl Phys. Lett.} {\bf
B297} (1992) 111.}
\REF\clineraby{%
J. Cline and S. Raby, {\sl Phys. Rev.} {\bf D43} (1991) 1781.}
In low-energy supersymmetry, one typically expects the gravitino to
possess a mass of order $\msusy$.  The gravitino may or may not be the
lightest supersymmetric particle.\foot{I shall stick to the notation
in which the LSP is the lightest supersymmetric particle, excluding
the gravitino.}  Since its interactions with
ordinary matter are gravitational in strength, its lifetime would
exceed the lifetime of the universe by many orders of magnitude.
Thus, the gravitino is another candidate for dark matter (in addition
to the LSP mentioned in section 2.8).  This is problematical, since as
shown in ref.~\gravitino, a gravitino whose mass is of order
$\msusy\simeq 100$~GeV---1~TeV would lead to a mass density of the
universe significantly larger than the critical density.  That is, the
number of primordial gravitinos is predicted to be too large.
One solution to this problem is to suppose that the universe
reheats only up to about $10^{10}$~GeV after inflation.  Inflation
dilutes the primordial gravitinos sufficiently and a low reheating
temperature would insure that gravitinos are not regenerated in
significant numbers.\refmark\moregravitino\
In the early days of supersymmetry model
building, such a solution was disfavored, since
a successful model of baryogenesis at the
GUT scale implied that the baryons were generated after inflation,
which required a reheating temperature substantially above
$10^{10}$~GeV.  Recently, there has been much theoretical work which
indicates that baryogenesis at the electroweak scale\refmark\baryons\
is possible.\foot{In low-energy supersymmetric models, electroweak
baryogenesis is possible in the context of the MSSM.\refmark\nelson}
In this case, inflation and a low reheating temperature can be a
viable solution to the gravitino problem.  (See also
ref.~\clineraby\ for an alternative suggestion.)

\break
\item{7.} {\sl Flavor changing neutral current (FCNC) problems.}

\vskip2.5pt
One of the great successes of the Standard Model is that FCNCs are
very suppressed, as required by experimental bounds on FCNC
processes.  The suppression of FCNCs is a consequence of the GIM-%
mechanism.  On the other hand, the origin of flavor in the Standard
Model is a complete mystery.  Extended
technicolor models attempt to solve both the origin of electroweak
symmetry breaking and the origin of flavor
with new physics in the
energy range between 1~TeV and 1000~TeV.  Perhaps it is not
surprising that such an ambitious program is generally plagued
with FCNCs which violate the strict experimental bounds.

\REF\gabb{See, \eg,
F. Gabbiani and A. Masiero, {\sl Nucl. Phys.} {\bf B322}
(1989) 235.}
Supersymmetry is often touted as being superior in that there is no
FCNC problem.  This is only partially correct.  To avoid FCNCs,
it must be true that the squark mass matrices are approximately
diagonal in the same basis that the corresponding
quark mass matrices are diagonal.
In addition, since the dominant
contributions to the squark masses arise from soft-supersymmetry
breaking terms, one finds that the squarks must be roughly degenerate
in mass.\refmark\gabb
\foot{Phenomenological requirements impose strong constraints
only on the first two generations of squarks.  In the $\widetilde
q_L$--$\widetilde q_R$ basis, significant off-diagonal mixing in the
bottom and top-squarks sector (which splits the corresponding
squark masses from the common diagonal soft-supersymmetry-%
breaking mass) cannot be ruled out at present.}
In supergravity model building, a standard assumption is that all
soft-supersymmetry-breaking scalar masses at the Planck scale
are universal.  Of course, flavor information does enter the
renormalization group evolution, so that the low-energy squark mass
parameters will not be exactly flavor independent.  Nevertheless, it
is easy to show that the assumption of universal soft scalar masses
at the Planck scale is sufficient to keep FCNCs below their
experimental upper limits.

\REF\dilaton{V. Kaplunovsky and J. Louis, ref.~\stringsusy;
R. Barbieri, J. Louis and M. Moretti, CERN-TH.6856/93 (1993).}
\REF\munoz{B. de Carlos, J.A. Casas and C. Mu\~noz, {\sl Phys.
Lett.} {\bf B299} (1993) 234.}
\REF\seiberg{For a recent discussion, see Y. Nir and N. Seiberg,
{\sl Phys. Lett.} {\bf B309} (1993) 337.}
\REF\kagan{M. Dine, R.G. Leigh and
A. Kagan, SLAC-PUB-6147 (1993) and SCIPP 93/04 (1993).}
Are universal soft scalar mass terms at the Planck scale natural?
The answer appears to be model-dependent.  Such a result appears
automatically in supergravity models with canonical kinetic energy
terms, although there is no fundamental reason why a theory of
supergravity should only possess the simplest kinetic energy terms.
In superstring models (at string tree-level),
universal scalar masses appear in models in
which the supersymmetry-breaking arises solely from the
dilaton $F$-term.\refmark\dilaton\  More general models of
low-energy supergravity generate non-universal soft scalar masses,
although the corrections to universality is calculable and in some
cases may be sufficiently small.\refmark\munoz\
Another possibility, where the required squark degeneracy
is obtained by exploiting flavor symmetries at the Planck scale,
is explored in refs.~\seiberg\ and \kagan.

\REF\rabyhall{L.J. Hall, V.A. Kostelecky and S. Raby,
{\sl Nucl. Phys.} {\bf B267} (1986) 415.}
There are other dangers lurking if one begins to allow for new
physics at intermediate scales (between the TeV scale and the
Planck scale).  As shown in ref.~\rabyhall, integrating out the
effects of physics at an intermediate scale can produce effective
non-universal scalar mass terms at that scale.  Evolving the
parameters of the effective Lagrangian down to the electroweak scale
can generate FCNCs larger than the allowed bounds.
This is an important constraint on models that attempt
to attribute the origin of flavor to an intermediate scale.

\goodbreak
\item{8.} {\sl The flavor puzzle.}

\vskip2.5pt
\REF\dinenelson{M. Dine and A.E. Nelson, SCIPP 93/03 (1993).}
\REF\fmatrices{G. Anderson, S. Raby, S. Dimopoulos, L.J. Hall and
G. Starkman, LBL-33531 (1993).}
\REF\shafi{B. Ananthanarayan, G. Lazarides and Q. Shafi,
{\sl Phys. Rev.} {\bf D44} (1991) 1613; {\sl
Phys. Lett.} {\bf B300} (1993) 245; L.J. Hall, R. Rattazzi and U. Sarid,
LBL-33997 (1993).}
As mentioned in the previous item, the Standard Model and the MSSM
treat the fermion generations in the same way.  Neither provide any
insight into the origin of quark and lepton masses and mixing angles.
The discussion of section 3.7 suggests that in a fundamental
supersymmetric model, the origin of flavor probably lies at the
Planck scale.\footnote{\star}{Flavor could arise from
intermediate scale physics.\refmark\dinenelson\
But realistic models of this kind
(that satisfy, \eg, the FCNC constraints) are difficult to construct.
In addition, the presence of intermediate scales could seriously
disrupt the successful unification of gauge coupling constants
discussed in section 2.5.}
At least two different scenarios are possible.  In the first
scenario, the physics of flavor is imprinted on the fermion-Higgs
Yukawa couplings that arise from the underlying superstring theory.
Examples are known in which the Yukawa couplings are computable and
depend on topological properties of the compactified space that
defines the string vacuum.\refmark{\witten,\yukawas}
In the second scenario, quark and lepton
mass matrices are generated from the dynamics at the grand unified
scale.  Examples of this approach have recently appeared in
refs.~\fermimass\ and
\fmatrices.  In this approach, supersymmetry does not play a
fundamental role in the generation of the quark and lepton matrices;
rather it is required in order to have
consistent unification of couplings.
One particularly elegant scenario suggests that the {\it three}
third-generation Yukawa couplings ($h_t$, $h_b$ and
$h_\tau$) all unify at some large scale $M_X$.\refmark{\fmatrices,\shafi}
Such models
predict that $\mt\sim 180$~GeV and $\tanb\sim m_t/m_b$.  Large $\tanb$
models have a number of interesting phenomenological implications
including enhanced radiative corrections to the light Higgs mass (if
$\mha>\mhl$) and enhanced Higgs couplings to the $b$-quark (and
$\tau$-lepton).

\item{9.} {\sl The CP-violation puzzle.}

\vskip2.5pt
\REF\phases{For example, see M. Dugan, B. Grinstein, and L.J. Hall,
{\sl Nucl. Phys.} {\bf B255} (1985) 413.}
\REF\cpgauge{M. Dine, R.G. Leigh and D.A. MacIntire, {\sl Phys.
Rev. Lett.} {\bf 69} (1992) 2030;
K. Choi, D.B. Kaplan and A.E. Nelson, {\sl Nucl.
Phys.} {\bf B391} (1993) 515.}
\REF\habernir{H.E. Haber and Y. Nir, {\sl Nucl. Phys.} {\bf B335} (1990)
363; L. Bento and G.C. Branco, {\sl Phys. Lett.} {\bf B245} (1990) 599.}
\REF\nelsonbarr{A. Nelson, {\sl Phys. Lett.} {\bf 136B} (1984) 387;
S.M. Barr, {\sl Phys. Rev. Lett.} {\bf 53} (1984) 329.}
\REF\sponcp{S.M. Barr and A. Masiero, {\sl Phys. Rev.} {\bf D38}
(1988) 366; S.M. Barr and G. Segre, {\sl Phys. Rev.} {\bf D48} (1993)
302; M. Dine, R.G. Leigh and A. Kagan, SLAC-PUB-6090 and SCIPP 93/05
(1993), {\sl Phys. Rev.} {\bf D49} (1993) in press.}
\REF\dannen{A. Dannenberg, L. Hall and L. Randall, {\sl Nucl. Phys.}
{\bf B271} (1986) 574.}
\REF\pomarol{A. Pomarol, {\sl Phys. Lett.} {\bf B287} (1992) 331;
{\sl Phys. Rev.} {\bf D47} (1992) 273.}
In the Standard Model, CP-violation arises
{}from a complex phase of the CKM-matrix.  This complex phase is also
a source of CP-violation in the MSSM; in both cases,
there is no clue to the fundamental origin of CP-violation, or the
relevant energy scale involved.  The solution to this problem may be
intimately connected to the flavor puzzle discussed above, since the
CKM-phase arises after diagonalizing the quark mass matrix.  However,
supersymmetric theories introduce new
complex phases.  For example, in the MSSM
complex phases can appear in the gaugino Majorana
mass terms and the $A$-parameters.\refmark\phases\
If these phases were ${\cal O}(1)$,
one would compute an electric dipole moment for the neutron
which is 2 to 3 orders of magnitude larger than the present
experimental bounds.  Thus, one must conclude that the new
supersymmetric phases are no larger that $10^{-3}$--$10^{-2}$.  Can
such a result arise in a natural way?

In the MSSM, one typically sets the unwanted phases to zero.  But in
a more fundamental supersymmetric theory, the question of these
phases must be addressed.  If these phases are zero at the Planck
scale, then their magnitudes at the electroweak scale (driven by
renormalization group evolution) are certainly small enough to avoid
potential phenomenological problems.  But, this then shifts the
question to the Planck scale.  What sets the Planck scale phases to
zero?

String theory may provide a hint at a solution to this problem.  It
turns out that in string theory, the CP-transformation is in fact a
gauge transformation in higher-dimensional space-time.\refmark\cpgauge\
Since gauge symmetries cannot be
explicitly broken, it follows that at a fundamental level, CP-%
violation must arise from a spontaneous breaking of CP at some higher
energy scale (perhaps the Planck scale).  This could provide a
theoretical motivation for setting Planck scale phases to zero.
Remarkably, in a model of spontaneous CP-breaking at a very high
scale, upon integrating out the physics at the high scale, the
effective low-energy CP-violation has precisely the form of a single
phase in the CKM-matrix.\refmark{\habernir-\sponcp}
This scenario requires new intermediate
scale (or Planck scale) physics associated with the scale of
CP-violation.\refmark{\nelsonbarr-\dannen}
It appears impossible to construct a viable model of
spontaneous CP-violation solely
in the context of the MSSM.\refmark\pomarol\

\REF\smalltheta{J. Ellis and M.K. Gaillard, {\sl  Nucl. Phys.}
{\bf B150} (1979) 141.}
\REF\thetaqcd{R. Akhouri, I. Bigi and H.E. Haber, {\sl Phys.
Lett.} {\bf 135B} (1984) 113; R. Akhouri and I. Bigi, {\sl Nucl.
Phys.} {\bf B234} (1984) 459.}
Although the above scenario sounds compelling, there is a
potential problem associated with the strong CP phase.
Based on the present limits on the electric dipole moment of the
neutron, it is known that $\bar\theta<10^{-9}$.  But in models of
spontaneously broken CP, $\bar\theta$ is calculable.  One can think
of the mechanism as follows: set $\bar\theta=0$ at the high scale
where CP is a good symmetry.  Below the scale of CP-breaking,
a non-zero (finite) value of $\bar\theta$ is generated.
If only Standard Model particles remained after integrating out the
physics above the CP-breaking scale, one would find an incredibly
small value for $\bar\theta$ well below the experimental
limits.\refmark\smalltheta\
This result follows because one needs to go to a high order in
perturbation theory before the first nontrivial correction to
$\bar\theta$ arises.  But, in models with low-energy supersymmetry,
contributions to $\bar\theta$ may arise at one-loop and yield a value
for $\bar\theta$ larger than the experimental
bound.\refmark{\thetaqcd,\sponcp}
Thus, it is a
challenge to model-builders to construct a viable
supersymmetric model of spontaneously broken CP-violation which does
not generate a value of $\bar\theta$ that is incompatible with the
bound on the neutron electric dipole moment.

\item{10.} {\sl The origin of low-energy discrete symmetries.}
\vskip2.5pt
\REF\bcons{S. Weinberg, {\sl Phys. Rev. Lett.} {\bf 43} (1979)
1566; F. Wilczek, {\sl Phys. Rev. Lett.} {\bf 43} (1979) 1571.}
\REF\bconssusy{S. Weinberg, {\sl Phys. Rev.} {\bf D26} (1982)
287.}
One of the great triumphs of the Standard Model is that
SU(3)$\times$SU(2)$\times$U(1) gauge invariance is sufficient to
eliminate the possibility of baryon number ($B$) and lepton number
($L$) violating operators of dimension four or
less.\refmark\bcons\  This provides a
natural explanation why the Standard Model conserves $B$ and
$L$ to such great accuracy.\foot{In fact, $B$ and $L$ are not exact in
the Standard Model but are violated due to the electroweak anomaly.
But the size of such violations is exponentially suppressed and not
relevant to the discussion here.}
Unfortunately, this elegant result of the Standard Model is lost in
the MSSM.\refmark\bconssusy\  To see why, recall that the
supersymmetric interactions are fixed once one specifies
the superpotential.
The most general gauge-invariant superpotential of the MSSM
has the following form:
$$
  W = W_R + W_{NR}\,.\eqn\generalw
$$
First, $W_R$ is given by
$$
  W_R = \epsilon_{ij} \left[ h_\tau \widehat H^i_1 \widehat L^j
         \widehat E + h_b\widehat H^i_1 \widehat Q^j\widehat D
         - h_t \widehat H^i_2 \widehat Q^j\widehat U
         - \mu \widehat H^i_1 \widehat H^j_2 \right]\,.\eqn\wrparity
$$
In eq.~\wrparity, $\epsilon_{ij}$ is used to combine two SU(2)
doublets [where $\epsilon_{ij}=-\epsilon_{ji}$ with $\epsilon_{12}=1$].
The parameters introduced above are the Yukawa coupling matrices
$h_\tau$, $h_b$ and $h_t$ (generation labels are suppressed)
and the Higgs superfield mass parameter, $\mu$.
Second, $W_{NR}$ is given by
$$
  W_{NR} = \epsilon_{ij} \left[ \lambda_L \wh L^i\wh L^j \wh E +
           \lambda\pri_L \wh L^i \wh Q^j\wh D -
           \mu\pri \wh L^i\wh H^j_2 \right] + \lambda_B \wh U\wh D\wh D
\,,\eqn\norparity
$$
where generation labels are again suppressed.
One quickly observes that the terms in $W_{NR}$ violate either
baryon number (B) or lepton number (L).  Specifically,
$$ \eqalign{%
  \wh L\wh L\wh E,\ &\wh L\wh Q\wh D,\ \wh L\wh H\qquad\qquad
        \Delta L\neq 0 \,,\cr
   &\wh U\wh D\wh D   \qquad\qquad\qquad\,   \Delta B\neq 0 \,.\cr}
\eqn\blviolation
$$

In the MSSM, one sets $W_{NR} = 0$ in order to recover $B$ and $L$
symmetry.  This can be implemented by introducing
a discrete symmetry.  There are two equivalent descriptions:
\REF\matterparity{S. Dimopoulos, S. Raby and F. Wilczek,
{\sl Phys. Lett.} {\bf 112B} (1982) 133.}
\REF\fayet{P. Fayet, {\sl Nucl. Phys.} {\bf B90} (1975) 104; {\sl
Phys. Lett.} {\bf 69B} (1977) 489.}
\REF\weirdrparity{L.J. Hall, {\sl Mod. Phys. Lett.} {\bf A5} (1990) 467.}
\REF\hallrandalltwo{%
L.J. Hall and L. Randall, {\sl Nucl. Phys.} {\bf B352} (1991) 289.}
\REF\massivegluinos{P. Fayet, {\sl Phys. Lett.} {\bf 78B} (1978) 417.}
\item{(i)}
Matter parity\refmark\matterparity\

\noindent
The MSSM does not distinguish between Higgs and quark/lepton superfields.
One can define a discrete matter parity under which all quark/lepton
superfields are odd while the Higgs superfields are even.

\goodbreak
\item {(ii)}  R-parity\refmark\fayet\

\noindent
\REF\wessbagger{J. Wess and J. Bagger, {\it Supersymmetry and
Supergravity} (Princeton University Press, Princeton, NJ, 1992).}
In the supersymmetric limit, one can show that the theory possesses
a continuous U(1)$_R$ symmetry if $R=2$ for all terms in the
superpotential $W$.\foot{My normalization of the $R$-quantum
number differs by a factor of two from that of
ref.~\wessbagger.}
It follows that in order to set $W_{NR}=0$,
one may choose
$$\eqalign{%
  R&= 1 \qquad\qquad  \hbox{for}\; \wh H_1,\wh H_2\,, \cr
  R&= \half\qquad\qquad
  \hbox{for}\; \wh L, \wh E,\wh Q,\wh U,\wh D\,. \cr }
\eqn\rparitychoice
$$
The full continuous U(1)$_R$ symmetry does not survive when the
soft-supersymmetry-breaking terms are included;
the U(1)$_R$ symmetry breaks down to a discrete
$Z_2$ symmetry called $R$-parity.  It is easy to check that
the $R$-parity quantum number is given by
$$
  R = (-1)^{3(B-L) + 2S}\eqn\ztwo
$$
for particles of spin $S$.\foot{Interesting
alternative supersymmetric models exist in which the $R$-parity
symmetry described above is modified.  Among such models are
$R$-parity-violating models, and models that promote the $Z_2$ $R$-parity
of the MSSM to a larger discrete symmetry group\refmark\weirdrparity\
or even to the full
continuous U(1)$_R$ symmetry.\refmark\hallrandalltwo\
In the latter case, one must introduce
new color octet fermions to mix with the gluinos,\refmark\massivegluinos\
in which case
U(1)$_R$-symmetric massive color-octet Majorana fermions are permitted.
Such models represent interesting
alternatives to the MSSM.}

\REF\merlo{S. Dimopoulos and L.J. Hall, {\sl Phys. Lett.} {\bf B207}
(1987) 210;
S. Dimopoulos, R. Esmailzadeh, L.J. Hall, J.-P. Merlo and G.D. Starkman,
{\sl Phys. Rev.} {\bf D41} (1990) 2099.}
\REF\herbie{H. Dreiner and G.G. Ross,
{\sl Nucl. Phys.} {\bf B365} (1991) 597.}
If low-energy supersymmetry is correct, will it be $R$-parity
invariant?  From a purely phenomenological point of view, one cannot
rule out the possibility that some of the operators listed in
eq.~\blviolation\ are present.
One can set bounds on $\lambda_L,\lambda'_L, \mu'$ and $\lambda_B$
[see eq.~\norparity],
based on $B$ and $L$ violation limits in a variety of Standard Model
processes.\refmark{\merlo,\herbie}  There is one weak argument in favor
of a conserved $R$-parity.  If $R$-parity is violated, then the LSP is
no longer stable and therefore cannot be the dark matter.  If one
regards the existence of dark matter as one of the selling points for
low-energy supersymmetry (see section 2.8), then
one must demand that $R$-parity is a good symmetry.

\REF\discrete{B.R. Green, K.H. Kirklin, P.J. Miron and G.G. Ross,
{\sl Nucl. Phys.} {\bf B278} (1986) 667; {\bf B292} (1987) 606;
R. Arnowitt and P. Nath, {\sl Phys. Rev. Lett.} {\bf 62} (1989)
2225; {\sl Phys. Rev.} {\bf D40} (1989) 191; {\bf D42} (1990)
2948.}
Whether one believes in $R$-parity or not, it is clear that
phenomenological requirements prevent the simultaneous appearance of
all the operators listed in eq.~\blviolation.  Thus, it seems
inevitable that some sort of discrete symmetry will be required to
remove the unwanted operators.  Any
fundamental theory of supersymmetry must address the origin of
such discrete symmetries.  Examples of
superstring models are known in which the discrete symmetries
necessary for $R$-parity invariance emerge
naturally.\refmark\discrete\
This may provide a clue as to the origin of discrete
symmetries in low-energy supersymmetry.

\chapter{Conclusions}

With the discovery of the $W$ and $Z$ gauge bosons in the early 1980s,
particle physics entered the era of electroweak symmetry breaking.
Nevertheless, the fundamental origin of electroweak symmetry breaking
is still unknown.  Theoretical arguments have made a convincing case
that the dynamics underlying electroweak symmetry breaking must be
associated with physics at the TeV scale.  Low-energy supersymmetry is
a leading candidate for this dynamics, for the reasons presented in
section 2.  However, these ten reasons can never carry the weight of
one reason---the discovery of supersymmetric particles at some future
collider.  It is in this regard that the next generation of
supercolliders---the LHC and SSC---are indispensable.  These machines
are designed to have the capability of determining whether low-energy
supersymmetry or some other dynamics is responsible for the masses of
the gauge bosons.  Without the supercolliders, progress at the
forefront of particle physics will stop cold.

In the meantime, the top-ten list of section 3 provides a useful menu
of theoretical problems that supersymmetric theorists must address.
Perhaps some progress will occur as we await the deliberations of the
politicians.  But, I suspect that major theoretical breakthroughs
will elude us until we have more experimental input.  It will be
instructive to see how the top-ten lists presented here are viewed ten
and twenty years from now.  If supersymmetry survives, its main
promise will be that it provides a window to the Planck scale.  May we
be so fortunate!

\vskip\chapterskip
\centerline{\bf Acknowledgements}
\vskip .1in

The invitation and hospitality of Jorge Lopez and his colleagues
is much appreciated.  I would like to thank Gordy Kane whose
advocacy of the phenomenological hints for supersymmetry inspired
this talk.
In addition, I gratefully
acknowledge conversations with Michael Dine, Robert
Leigh and Pierre Ramond.  Finally, I thank the hospitality
of the Aspen Center for Physics which provided me with another
opportunity to give this talk, and where the written version of
this work was completed.
This work was supported in part by the Department of Energy
and in part by the Texas National Research Laboratory Commission
grant \#RGFY93-330.

\refout
\endpage \figout
\bye